\documentclass[aps,pre%
,twocolumn%
,showpacs%
,superscriptaddress%
]{revtex4}

\usepackage{graphicx}
\usepackage{times}

\begin{document}

\newcommand{\affila}{
 Department of Biological Physics, E\"otv\"os University,
 P\'azm\'any P.\ stny.\ 1A, H-1117 Budapest, Hungary
}
\newcommand{\affilb}{
 Biological Physics Research Group of HAS,
 P\'azm\'any P.\ stny.\ 1A, H-1117 Budapest, Hungary
}

\title{
 Phase transition in the collective migration of tissue cells: experiment and
 model
}

\author{B. Szab\'o}
 \affiliation{\affila}
\author{G. J. Sz\" oll\H{o}si}
\author{B. G\"onci}
 \affiliation{\affila}
 \author{Zs. Jur\'anyi}
 \affiliation{\affila}
 \author{D. Selmeczi}
 \affiliation{\affila}
\author{Tam\'as Vicsek}
 \affiliation{\affila}
 \affiliation{\affilb}

\date[]{\protect\today}

\begin{abstract}
We have recorded the swarming-like collective migration of a
large number of keratocytes (tissue cells obtained from the scales
of goldfish) using long-term videomicroscopy. By increasing the
overall density of the migrating cells, we have been able to
demonstrate experimentally a kinetic phase transition from a
disordered into an ordered state. Near the critical density a
complex picture emerges with interacting clusters of cells moving
in groups. Motivated by these experiments we have constructed a flocking model
that exhibits a continuous transition to the ordered phase, while assuming
only short-range
interactions and no explicit information about the knowledge of the 
directions of motion of neighbors. Placing cells in microfabricated arenas
we found spectacular whirling behavior which we could also reproduce in simulations.
\end{abstract}


\maketitle
\section{Introduction}
The collective motion of organisms is a spectacular phenomenon
sometimes involving huge schools of fish, thousands of birds
exhibiting complex aerial displays \cite{parrish}, herds of
quadrupeds and even bacteria producing fractal colonies
\cite{czirokpre} or amoeba assembling into rotating
aggregates\cite{levine_1999}. 
Recently, sperm
cells were also demonstrated to form self-organized vortices 
\cite{collective_sperms}.
In addition to being a
common mechanism by which organisms self-organize, a deeper
understanding of the simultaneous adjustment of the velocities of
many moving objects has important potential applications ranging
from the swarming of distributed robots exploring new territories
\cite{jadbabaie} to the healing of wounds related to the coherent
migration of epithelial cells \cite{collective_epithelial_cells_on_collagen}.

Although widely observed in nature, collective motion is less
accessible for experimental investigations under laboratory
conditions. Becco {\em et al.\ } \cite{fish} have presented interesting
but only qualitative results. A well controlled series of experiments
aimed at characterizing the nature of the transition from a
disordered to an ordered phase in velocity space is so far
lacking, even though such experiments would
be useful both in providing a quantitative reference for
further experimental studies and in prompting more realistic models
for group behavior. While laboratory observations have been
scarce, a number of models based on self-propelled
particles have been developed recently
(see, e.g., Refs. \cite{vicsek,toner,gregoire,shimoyama,mikhailov,levine_2000,
erdmann}) to
describe collective motion. In broad terms these models fall into two 
categories, those which describe 
the onset of collective motion as a transition to an ordered
state in a large noisy system with simple interactions between the particles
\cite{vicsek,gregoire,toner}, 
and those which employ more complicated -- consequently more
realistic -- interactions and focus on the emergent collective 
dynamics of groups of finite
size \cite{shimoyama,mikhailov,levine_2000,erdmann}.  
A model, however, that combines simple short-ranged  interactions, 
detailed realistic dynamics and a well defined 
kinetic transition in a large noisy system has not been described so far.

In this paper we present experimental results concerning the
collective migration of a large number of tissue cells
(keratocytes) using long-term videomicroscopy. 
As the overall
density of the migrating cells is increased, we observe a kinetic
phase transition from a disordered (low density) state into an ordered
(high density) state, in which most of the cells move in a
direction approximately agreeing with their average direction of
motion. Just below the transition a complex picture emerges with
interacting groups of cells moving in random directions. 
Motivated by these experimental results we develop a flocking model,
which, in contrast to previous models, considers a minimal realistic
interaction that assumes no explicit averaging 
of the directions of motion, while also exhibiting a 
transition to the ordered phase. 
Numerical studies indicate that this transition is continuous
and belongs to the same universality class as
the model of Vicsek {\emph et al.} \cite{vicsek}. These results compel 
us to imply that the experimental
transition described is continuous as well.
\begin{figure*}
\centerline{
\mbox{\includegraphics[angle=0,width=0.75
\textwidth]{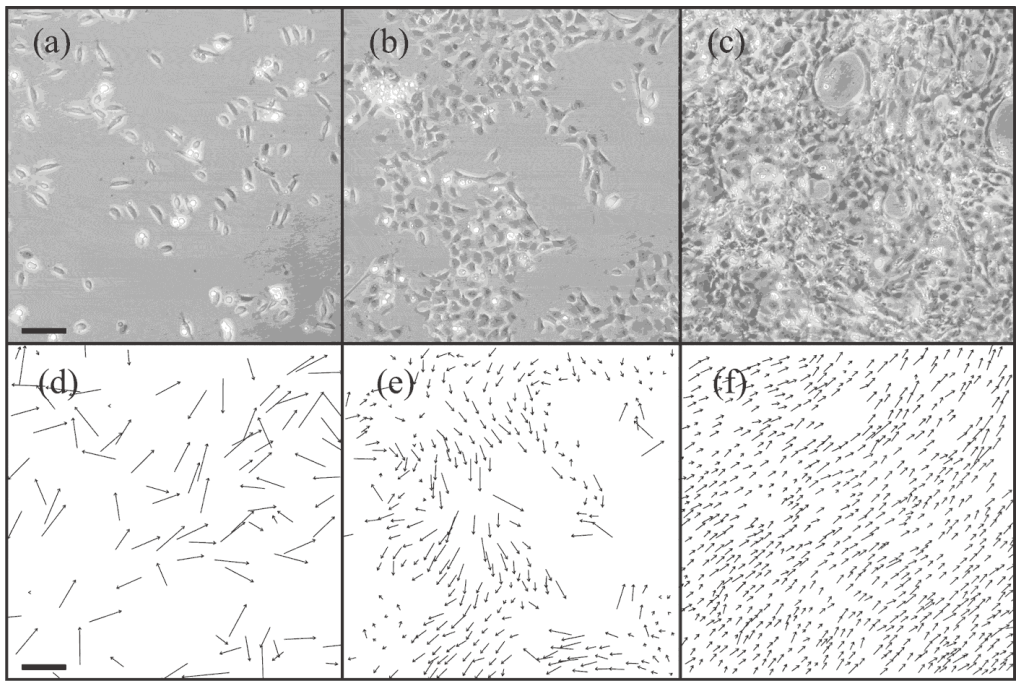}}}
\caption{Phase contrast images showing the typical behavior of cells for
three different densities. (a): 1.8, (b): 5.3, (c): 14.7 cells/100x100 
$\mu$m$^2$. 
We observed that as cell density increases cell motility undergoes collective
ordering. The speed of single cells is higher than that of cells moving 
in coherent groups. Scale bar: 
200 $\mu$m. (d)-(f): Velocity of cells. Scale bar: 50 $\mu$m/min. 
Online  supplemental material 
\cite{online} contains corresponding videos.} 
\label{fig1}
\end{figure*}

\section{Experimental Setup and Results}
Our experimental setup, consisting of a home developed 
fully computer-controlled 
time-lapse microscope \cite{jojo} and a custom made 
room temperature ${\rm CO}_2$ mini-incubator 
 allowed us to carry 
out long-term videomicroscopy of keratocytes together with a
quantitative analysis of their motion.
We collected 2--4 fish scales from living goldfish 
({\em Carassius auratus}) with tweezers, and placed them external 
side up in a 35-mm circular Petri dish similarly to Ref. \cite{Borisy_keratocytes}. 
Scales were kept in the incubator
overnight 
to allow epidermal keratocyte cells to migrate out from the scales. 
Before time-lapse microscopy scales 
were removed, and the cells remaining in the Petri dish were treated 
with phosphate saline buffer and/or trypsin to obtain cultures of varying density 
(both compounds reversibly weaken cell-cell and cell-substrate connections, 
allowing 
the removal 
of  a controllable percentage of the cells).
During the subsequent typically 24--h long time-lapse 
microscopy experiments we monitored the motion of live keratocyte 
cells (taking pictures in several fields of view at frequency of 1 shot per minute) in 
cultures of varying density ranging from sparse ones with very low cell densities 
to confluent ones with nearly complete coverage.  We observed a relatively 
sharp transition from random motility to an ordered collective migration of dense 
islands of cells as the density was increased. 
Fig.\ \ref{fig1} shows the typical behavior of cells 
for three different densities. The technical details of the experiment are 
described in the online supplemental material \cite{online}. 

\begin{figure}
\centerline{
\includegraphics[
width=1.1 
\columnwidth]{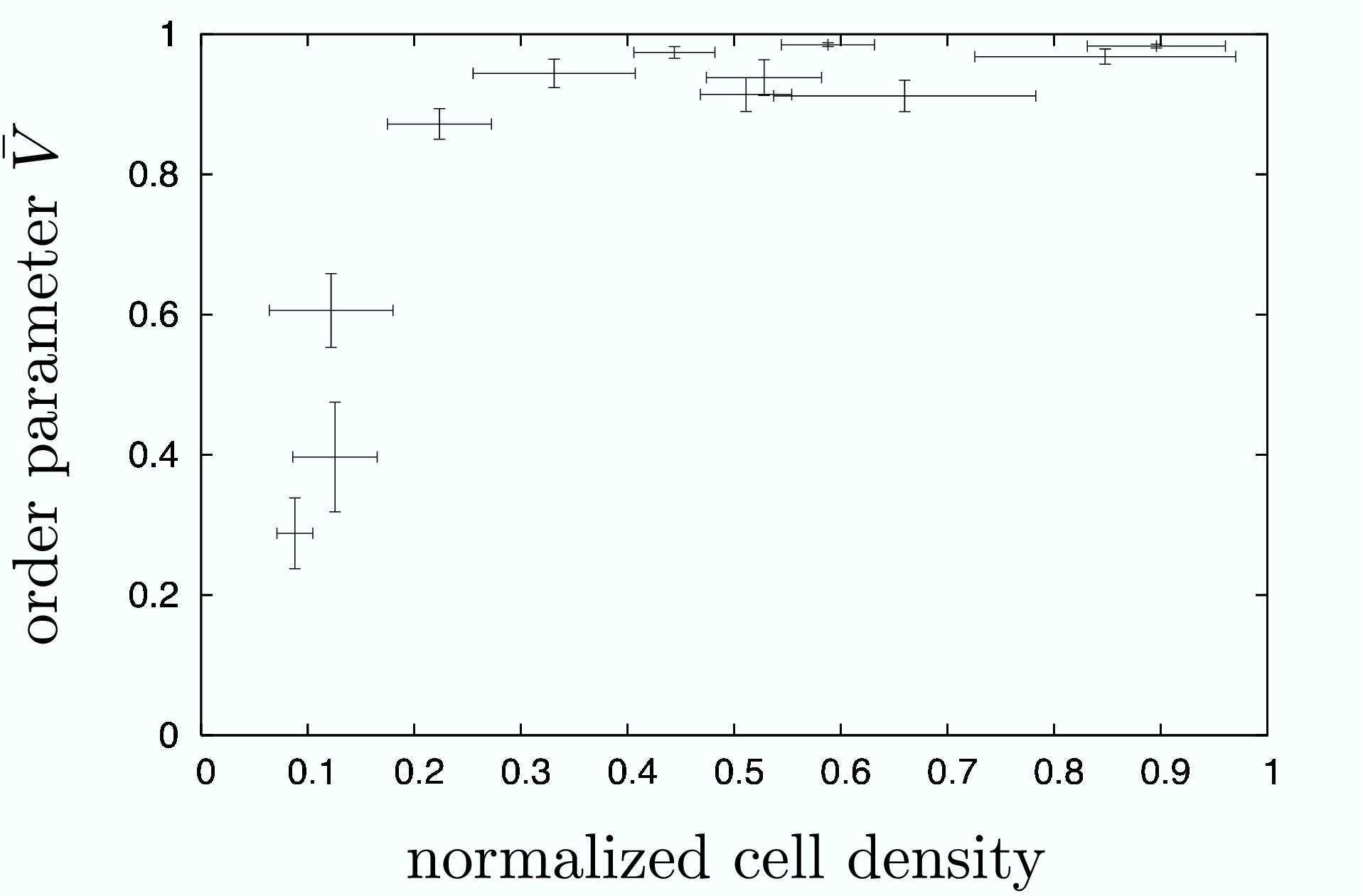}}
\caption{Order parameter $\bar V$ is shown as a function of normalized 
cell density. 
 Cell density was normalized 
with the maximal observed density of 2.5 $\times$ 10$^{-3}$
cells/$\mu$m$^2$
and error bars indicate the standard error of the density and order parameter.}
\label{fig2}
\end{figure}

\begin{figure*}
\centerline{\includegraphics[angle=0,width=2.
\columnwidth]{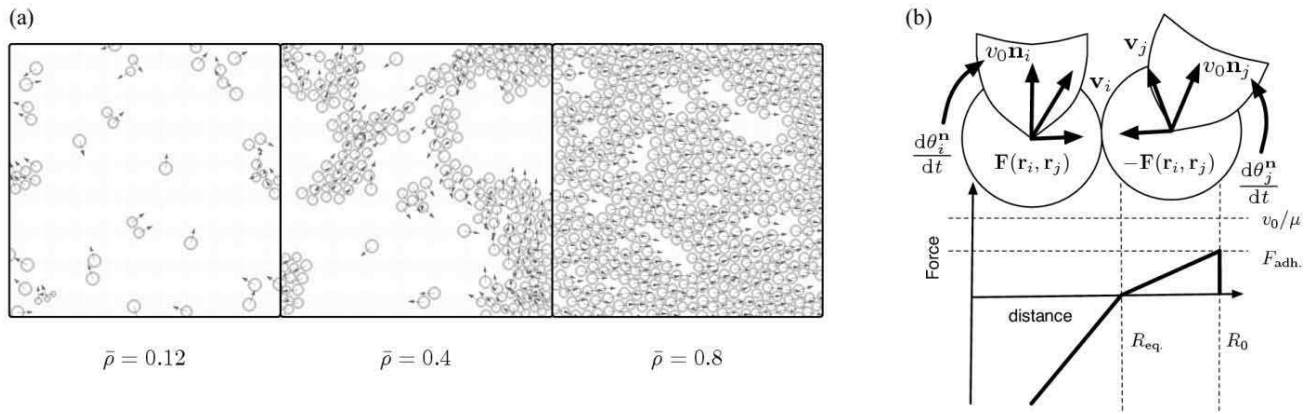}}
\caption{Computer simulations were performed of
 the system described in the text for different densities.
 These simulations showed a transition to the ordered phase similar 
to that seen in experiments. 
 (a) Typical behavior of cells is shown for three different 
values 
of the normalized number density $\bar\rho= \rho/\rho_{\rm max}$, 
with $\rho_{\mathrm{max}} \approx 2$,  
which is approximately the density where
 gaps disappear,
  and the cells reach tight packing in simulations. 
(b) As cells moving in different directions come into contact
adhesive
  inter-cellular forces act to align  ${\bf n}_i$ and ${\bf n}_j$, resulting
  in an effective averaging of self-driving directions. For videos of
  simulation runs see the online supplemental material \cite{online}.
}
\label{modelfig}
\end{figure*} 

In order to quantify  the level of coherence 
of the migration pattern of cells --the individual motion of which 
are usually described as a persistent random 
walk \cite{Selmeczi}-- we calculated the velocity of 20-30 cells 
in each experiment from their displacement 
between frames. This process yielded the ${\bf v}_{i}(t_{k})$ 
velocity vectors of cells, where $i$ is the index of the cell,
$t_{k}=k \Delta t$ is the time elapsed from the start of the
cell's trajectory and $\Delta t$ denotes the time difference
between frames.  We define as the measure of coherence of motion 
an order parameter equal to the 
time average of the sum of the
normalized velocities divided by the number of cells measured. 
Thus, the order parameter is  
\begin{equation}
 \bar V = \left\langle \frac{1}{N} \left|
\sum_{i=1}^N \frac{{\bf v}_i(t_k)}{|{\bf v}_i(t_k)|}
\right| \right\rangle_{t_k},
\label{experimental_order_parameter}
\end{equation}
where $N$ is the
number of evaluated cells. 
Fig.\  \ref{fig2} displays 
the order parameter as a function of cell density.
Our measurements were carried out after the cells had had sufficient time
to migrate out of the scales and achieve a quasi-stationary migration.
$\bar V$ 
was calculated by averaging over the observation time and its standard error was calculated by dividing the standard deviation 
by $\sqrt{M-1}$, where $M$ is the number of analyzed snapshots. 
We attempted to survey the extent of finite size
effects, by calculating the order parameter 
and its error, while considering only half the number of available
cells. We found no significant deviation. Cell density was measured
locally, i.e.\ in the 
field of view, every 30 minutes. Both its mean value and the standard 
error of the mean is presented in Fig.\ 2.
A sharp increase can be observed around 5 $\times$ 10$^{-4}$ 
cells/$\mu$m$^2$. 

We conclude that a kinetic phase transition takes place from a 
disordered into an ordered state as cell density exceeds a 
relatively well-defined critical value. Our experiments suggest that 
short-range attractive--repulsive intercellular forces alone are
sufficient to organize motile keratocyte cells into coherent 
groups.
\section{Model Description and Results}

To interpret the above phase transition-like ordering
phenomenon we constructed a model that takes into account the specific
features of the experimental system and is able to reproduce 
experiential behavior. 
In our model individual model cells (self-propelled particles)
move forward in a well defined direction with 
constant speeds. The noisy nature of the processes which generate 
cell locomotion is taken into account by considering the direction of
self-propulsion of model cells to be noisy. Intercellular forces through which
model cells
interact are considered to be short-ranged, 
as between keratocytes they are the result of
direct physical contact. Further, regarding interactions 
between keratocyte cells, it is obvious that explicit 
averaging of the directions of motion employed in previous models
is not realistic. Tissue cells forming
coherently migrating groups are unable to explicitly
adjust their direction of motion to the average velocity of their
neighbors, collective motion must emerge solely as a result of direct
cell to cell interactions (forces). To model the emergence of
collective motion without such explicit averaging, we consider
self-propelled particles (model cells) that attempt to adjust their
direction of motion toward the direction of the net-force acting on
them. 

The 2-dimensional motion of  model cell  $i \in \{1,N\}$ with position
${\bf r}_i (t)$ is described by the overdamped dynamics: 
\begin{equation}
 \frac{{\rm d }{\bf r}_i (t)}{{\rm d} t } = v_0 {\bf n}_i (t)
 + \mu \sum_{j=1}^{N} {\bf F}({\bf r}_i,{\bf r}_j).
\label{eomv}
\end{equation}
Thus, each cell with mobility $\mu$ attempts to maintain a self-propelling velocity of magnitude $v_0$ in
the direction of the unit vector ${\bf n}_i (t)$ and experiences intercellular
forces ${\bf F}({\bf r}_i,{\bf r}_j)$.
The direction of the self-propelling velocity ${\bf n }_i (t)$, described by the
angle  $\theta^{\bf n}_i (t)$, attempts to relax to ${\bf v}_i(t)={\rm d}
{\bf r}_i (t) / {\rm d} t  $ with a relaxation time 
$\tau$, while also experiencing angular 
noise $\xi$:  
\begin{equation}
\frac{\mathrm{d} \theta^{\bf n}_i (t)}{\mathrm{d} t}  = 
\frac{1}{\tau} \arcsin( \left({\bf n}_i (t) \times
\frac{{\bf v}_i (t)}{|{\bf v}_i (t)|}\right) \cdot {\bf e}_z) +\xi,
 \label{eomn}
\end{equation}
where ${\bf e}_z$ is a unit vector orthogonal to the plane of motion
(see Fig.\ \ref{modelfig}b) and $\xi$ is a delta correlated
 Gaussian white noise term
with zero mean, ie.\ $\langle\xi(t)\rangle =0$ and  $\langle\xi(t)\xi(t')\rangle = \eta^2/12 \enskip \delta(t,t')$.
We consider pairwise inter-cellular forces whose magnitude is a function of the 
distance  $d_{ij}$  between
the two centers of mass only. 
Aiming for simplicity we present results obtained by using a 
piecewise linear force function, the existence of the transition we describe, 
however, does not depend on the specific form of the function employed. 
The piecewise linear force function we considered 
was repulsive for distances smaller then $R_{\rm eq. }$, 
attractive for distances $R_{\rm eq.} \leq d_{ij} \leq R_0$  and 
zero for cells farther apart, i.e.\ 
\begin{equation}
{\bf F}({\bf r}_i,{\bf r}_j) = {\bf e}_{ij} \times
\left\{\begin{array}{cl}
	F_{\rm rep.} \frac{ d_{ij} - R_{\rm eq.} }{R_{\rm eq.} }  
        &,\ d_{ij}<R_{\rm eq.} \\
	F_{\rm adh.} \frac{ d_{ij} - R_{\rm eq.} }{R_0-R_{\rm eq.} } 
        &,\ R_{\rm eq.} \leq d_{ij} \leq R_0\\
        0\enskip &,\ R_0<d_{ij}
	   \end{array}\right.,
\label{forcefunction}
\end{equation}
where ${\bf e}_{ij}= \frac{{\bf r}_i-{\bf r}_j}{| {\bf r}_i-{\bf r}_j
|}$, $d_{i j} = | {\bf r}_i-{\bf r}_j |$, $F_{\rm rep.}$ is the value
of the maximum repulsive force at $d_{ij}=0$ and $F_{\rm adh. }$ is
the maximum attractive force (resulting from adhesive interactions
between cells) at $d_{ij}=R_0$ (see Fig.\ \ref{modelfig}b). The values
$v_0=1$, $\mu=1$, $\tau=1$, $R_0=1$, $R_{\rm eq.} = 5/6 $, $F_{\rm
adh.}= 0.75 $ and $F_{\rm rep.}= 30 $ were used, the results obtained,
however, were not sensitive to the particular choice for the parameter
values. The parameters of the function (\ref{forcefunction}), the slope of the two linear segments ($F_{\rm rep.}$ and $F_{\rm adh.}$)
and the equilibrium distance $R_{\rm eq.}$, were
adjusted while observing the simulations and comparing them with
experimental videos (cf.\ online supplemental material), in order to achevie best possible agreement with the experiments. The two main criteria were:
i.) the reproduction of the
2-dimensional sheet-like motion, wherein individual
cells only interact with their nearest neighbors -- i.e.\ cells have a
well defined volume and ii) the ability of cells to
``break free'' from each other. 
 The first was readily achievable by setting the slope
of the repulsive force segment to be sufficiently larger than the that
of the attractive segment and the value of $R_{\rm eq.}$ sufficiently
larger than $R_0$, while the second necessitated that the
maximal value of the attractive segment $F_{\rm adh.}$ be set smaller
then $v_0/\mu$.
\begin{figure}
\includegraphics[angle=0,width= 1.
\columnwidth]{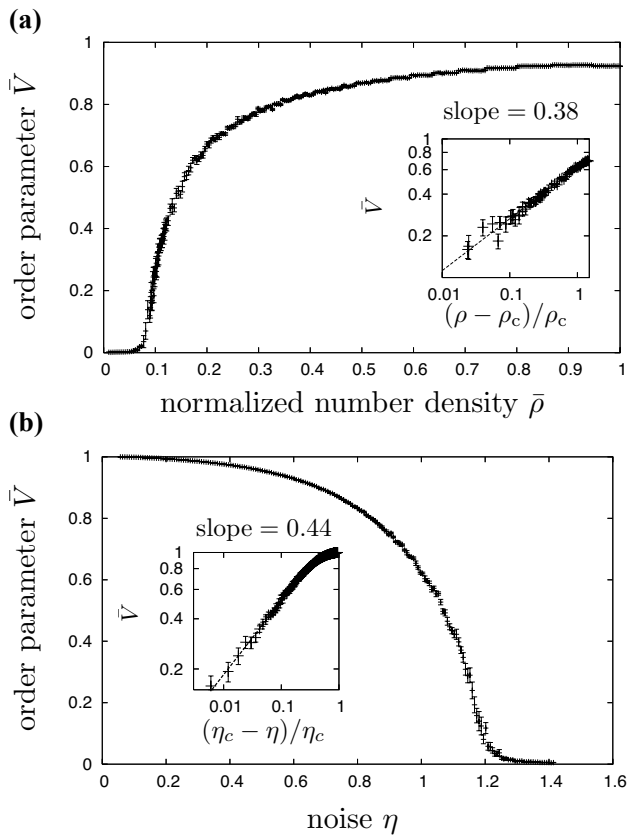}
 \caption{  The average value and standard error of $\bar V$ is shown 
as a function of the normalized number
  density (a) $\bar\rho= \rho/\rho_{\rm max}$ and (b) noise $\eta$. Each data point was obtained from at least $10$ 
independent simulation runs with $N=1000$, with (a) $\eta
=0.6$, $\rho_{\mathrm{max}} \approx2$ and (b) $\rho=0.6$. 
 The insets shows the dependence
  of $\ln \bar V$ on (a) $(\rho-\rho_{\rm c}(\eta))/\rho_c$ and (b) 
$(\eta_{\rm c}(\rho)-\eta)/\eta_c$,
  the slope of the fitted lines can be associated with the critical exponents
   $\delta$ and $\beta$ . 
The large scaling regimes and the similarity of the numerical
   values obtained for the exponents with those found for other models
indicate
    the existence of a continuous phase transition in both $\rho$ and $\eta$.
} \label{modelfig2}
\end{figure}

\begin{figure*}
\includegraphics[angle=0,width= 1.8
\columnwidth]{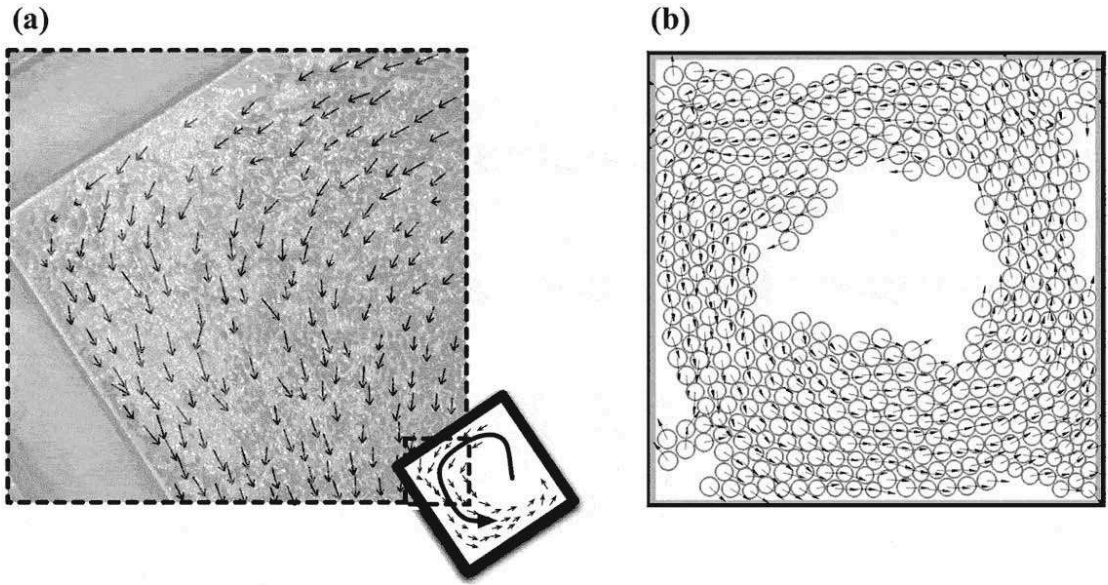}
 \caption{  (a) Experimental snapshot of a circulating cell group with
   the velocities of cells in the corner of a $2 \times 2$ mm$^2$ square shaped
    microfabricated arena. (b) Simulations of model cells confined to a
   square arena show the emergence of circular motion over a wide
   range of model parameters.
Supplemental video showing the circular
   motion of cells in the whole arena as well as in simulations are available online \cite{online}.
} \label{fig5}
\end{figure*}

We carried out simulations (Fig.\ \ref{modelfig}a) by solving the system of $3N$ 
differential equations (\ref{eomv}) and (\ref{eomn}) with periodic boundary
conditions in systems of size $L \times L$ using a
fixed time step of $\Delta t = 0.05 \enskip R_0/v_0$.
The angular noise term $\xi$ was modeled by choosing its magnitude uniformly from the
interval $ [-\eta/( 2 \sqrt \Delta t ) , \eta/( 2 \sqrt \Delta t )]$.
In good agreement with experiments, running our model resulted in
a continuous transition to the ordered phase (where all cells move
in a common direction) as the number density of cells $\rho=N/(L/R_0)^2$ 
was increased
for a fixed value of the noise $\eta=0.6$ (Fig.\ \ref{modelfig2}).
While varying the value of $\eta$ was not experimentally feasible,
simulations show that a continuous transition to the ordered phase
also takes place as $\eta$ is decreased while $\rho$ is held fixed
(data not shown, see online supplemental material \cite{online}). In both cases
we calculated the order parameter (\ref{experimental_order_parameter}) using  
the velocities ${\bf v}_i(t_k)$ obtained from the numerical solution of  
(\ref{eomv}) and (\ref{eomn}).
$\bar V$  and its standard error were calculated by averaging over at
least $10$
independent simulation runs each at least an
order of magnitude longer than the typical velocity autocorrelation time
(which was found to be in the range of $10-1000 \enskip R_0/v_0$). 
 The values  $\rho_{\rm c}=0.18\enskip (\bar\rho_{\rm c}=0.09)$ and 
$\eta_{c}=1.18$ 
were found 
for the critical density and noise from simulations of
systems with $N=250,500, 1000$ and $2000$ at fixed values of  $\eta=0.6$ and $\rho=0.3$ respectively. 
Based on the assumption that our model exhibits a kinetic phase
transition in the thermodynamic limit analogous to the continuous
phase transition in equilibrium systems, we proceeded to study
it's critical behavior,  that is,
\begin{equation}
\bar V \propto [\eta_{\rm c}(\rho)-\eta]^\beta \quad \quad \mathrm{and}
\quad \quad \bar V \propto  [\rho-\rho_{\rm c}(\eta)]^\delta,
\end{equation}
where $\beta$ and $\delta$ are the critical exponents and
$\eta_{\rm c}(\rho)$ and $\rho_{\rm c}(\eta)$ are the critical noise and
density. Analysis of the data yielded the values $\beta= 0.44 \pm
 0.08$ and $\delta=0.38 \pm 0.07$, strongly suggesting that our
model belongs to the same universality class as the angular noise
model of Vicsek \emph{et al.\ }\cite{vicsek}, for which
$\beta=0.45 \pm 0.07$ and $\delta=0.35 \pm 0.06$. 
 
When two cells approach each other close
enough, equation (\ref{eomn}) leads to a gradual alignment 
of their direction
of motion.
Our model is in this sense
similar to other models of systems of self-driven particles
exhibiting emergent collective motion -- i.e.\ flocking --, 
with the very important difference that the
particles -- the cells -- do not directly use information on the
movement of others around them to determine their own movement.
Also, while several other models include self-propelling particles, which
interact through various forces (and the seminal work of Shimoyama
{\emph et al.} \cite{shimoyama} has alignment dynamics similar to
ours), they all rely on either long-range forces \cite{shimoyama,erdmann}, explicit
averaging \cite{vicsek} or both \cite{levine_2000}.  
The model
presented above, on the other hand, combines an experimentally
motivated minimal dynamics with short-range adhesive and repulsive
forces, while also displaying a continuous kinetic transition to the
ordered state.
\section{Effects of boundary conditions}
To investigate the effects of boundary conditions on collective cell
motility, we used square as well as more complex shaped
microfabricated arenas, which kept cells in a well defined
area. Microstructures were fabricated by UV-curing Norland optical
adhesive (NOA63) on the surface of glass cover slips using UV
lithography. The typical diameter of the structures was 2 mm with 0.8
mm high walls. In closed 2D square shaped arenas we observed the
roundabout motion of large cell groups. 
Fig.\ \ref{fig5}
shows an experimental snapshot of the circular motion with the 
instantaneous velocities of cells in the corner of a square shaped
arena. A spectacular sustained whirling motion of the cells can be seen in the 
corresponding supplemental video \cite{online}. 

We were also able to reproduce the above effects of boundary conditions on
cell motion in simulations similar to those described in the previous
section. Placing model cells in a square box with
repulsive walls implemented through adding a repulsive force with an
exponential falloff and finite a cutoff to the
right side of equation (\ref{eomn}) corresponding to each of the four walls 
\begin{equation}
 {\bf F}_w(d_{ i w}) = {\bf n}_{ w} \times
\left\{
\begin{array}{cl}
  -F_{\rm wall} \enskip {\rm exp}(- \frac{2 d_{ i w }}{ R_0} )   
  &,\ d_{i w}<R_{0} \\
  0\enskip 
  &,\ R_0<d_{i w }\end{array}\right.,
\end{equation}
where $d_{\rm iw}$ is the distance between cell $i \in \{1,N\}$
and any of the $w \in \{1, 4\}$ four walls with unit normal vector
${\bf n}_w$ and $F_{\rm wall}= 50$. In simulations cells
circulated in an organized fashion under a wide range of noise and
density values as well as for different system sizes. The online
supplemental material \cite{online} contains a simulation videos
showing the emergence of organized circular motion of model cells
confined to a square box.

\section{Discussion}
In summary, we have presented evidence that purely short-range
forces and simple experimentally motivated dynamics can be equivalent 
to an effective alignment term.
Drawing an analogy between our experimental 
and model results, we imply that the emergence 
of collective motion among 
keratocytes is an example of a continuous kinetic phase transition.
Our results are also relevant in the broader context of recent work by
Gr\'egorie and Chat\'e \cite{gregoirechate} 
questioning the continuous nature of the transition in the
angular noise model. Our present numerical results (exponents being
similar to the ones determined for the original Vicsek {\em et al.\ }model \cite{vicsek} )
support the view \cite{mate} that, in contrast with Ref.\ \cite{gregoirechate},  angular noise models define a
universality class with a corresponding continuous phase transition in 
the ordering of the velocities. We expect that our and similar experiments as well as the
quantitative model of the observations we have provided
for the collective motion of tissue cells will lead to a better
understanding of such vital phenomena as wound healing or
embryogenesis.

\section*{ Acknowledgments}
We are grateful to Dr. G\'abor Cs\'ucs for his instructive help in
setting up a fish keratocyte lab in our institute. We thank Dr.
Zsuzsanna K\"ornyei for her advice in cell culturing and Prof. P\'al Ormos for the construction of the microfabricated structures. This study
was supported by Hungarian Science Research Funds: NKFP
3A/0005/2002 and OTKA Nos. F49795, T049674 .


\begin{thebibliography}{99}

\bibitem{parrish}
J. K. Parrish, and W. H. Hapner (eds.) {\em Animal Groups in
Three Dimensions} (Cambridge University Press, New York, 1997).

\bibitem{czirokpre}
A. Czir\'ok, E. Ben-Jacob, I. Cohen and T. Vicsek:
 { Phys. Rev. E,} {\bf 54}, 1791 (1996)

 \bibitem{levine_1999}
W.-J. Rappel, A. Nicol, A. Sarkissian, H. Levine, and W. F. Loomis
{ Phys. Rev. Lett.} {\bf 83}, 1247 (1999)

\bibitem{collective_sperms}
 I. H. Riedel, K. Kruse, J. Howard, { Science},
 {\bf 309,} 300-3 (2005)

\bibitem{jadbabaie}
A. Jadbabaie, J. Lin, and A. S. Morse { IEEE Transactions on
Automatic Control}, {\bf 48} 988 (2003)

\bibitem{collective_epithelial_cells_on_collagen}
H. Haga, C. Irahara, R. Kobayashi, T. Nakagaski, K. Kawabata, 
{ Biophys.\ J.\ } {\bf 88,} 2250-56 (2005)

\bibitem{fish}
Ch. Becco, N. Vandewalle, J. Delcourt, P. Poncin,
{ Physica A } {\bf 367}, 487 (2006)

\bibitem{vicsek}
T. Vicsek, A. Czir\'ok, E. Ben-Jacob, I. Cohen and O. Shochet, 
{ Phys. Rev. Lett.} {\bf  75}, 1226 (1995)

\bibitem{toner}
J. Toner and Y. Tu, 
{ Phys. Rev. Lett.}  {\bf 75}, 4326 (1995) 


\bibitem{gregoire}
G. Gr\'egoire, H. Chat\'e, and Y. Tu { Physica D} {\bf  181}, 157
(2003)

\bibitem{shimoyama}
N. Shimoyama, K. Sugawara, T. Mizuguchi, Y. Hayakawa 
and M. Sano,
{ Phys. Rev. Lett.} {\bf  76}, 3870 (1996)

\bibitem{mikhailov}
A. S. Mikhailov and D. H. Zanette,
{ Phys. Rev. E} {\bf 60}, 4571 (1999)

 \bibitem{levine_2000}
 H. Levine, W.-J. Rappel and I. Cohen,
{ Phys. Rev. E} {\bf 63}, 017101 (2000)

\bibitem{erdmann}
U. Erdmann, W. Ebeling and A. S. Mikhailov,
{ Phys. Rev. E} {\bf 71}, 051904 (2005)

\bibitem{jojo} B. Szab\'o, Zs. K\"ornyei, J. Z\'ach, D. Selmeczi, 
G. Cs\'ucs,
 A. Czir\'ok and T. Vicsek, { Cell Motil.\ Cytoskel.\ } {\bf 59} (1), 38 -49 (2004)

\bibitem{Borisy_keratocytes} T. M. Svitkina, A. B. Verkhovsky, 
K. M. McQuade,
G. M. Borisy, { J.\ Cell Biol.\ } {\bf 139} (2), 397-415 (1997)

\bibitem{online}
Supplemental material including experimental and simulation videos 
as well as technical details of the experimental setup is available at: 
{\bf http://angel.elte.hu/\textasciitilde bszabo/collectivecells}

\bibitem{Selmeczi} D. Selmeczi, S. Mosler, P. H. Hagedorn, N. B. Larsen,
 H. Flyvbjerg, { Biophys.\ J.\  }{\bf 89} (2), 912-31 (2005)

\bibitem{gregoirechate}
G. Gr\'egoire and H. Chat\'e
{Phys. Rev. Lett.} {\bf 92}, 025702 (2004) 

\bibitem{mate}
M. Nagy, I. Daruka, T. Vicsek  
{Physica A} accepted for publication


\end{thebibliography}
\end{document}